\documentclass[10pt,journal,compsoc]{IEEEtran}

\makeatletter
\newif\if@restonecol
\makeatother

\usepackage[linesnumbered,ruled,vlined]{algorithm2e}
\usepackage{algpseudocode}
\usepackage{amsmath}
\usepackage{color,cite}
\usepackage{cuted}
\newtheorem{defn}{Definition}
\usepackage{verbatim}

\title{Find Another Me Across the World - \\ Large-scale Semantic Trajectory Analysis Using Spark}

\author{Chaoquan~Cai,~\IEEEmembership{}
        Dan~Lin,~\IEEEmembership{Senior Member,~IEEE}

\IEEEcompsocitemizethanks{\IEEEcompsocthanksitem Chaoquan Cai is with EECS Department at University of Missouri, Columbia,
MO 65201, USA.\protect\\
E-mail: cckxf@mail.missouri.edu
\IEEEcompsocthanksitem Dan  Lin is with EECS Department at University of Missouri, Columbia, MO 65201, USA.\protect\\E-mail: lindan@missouri.edu}
\thanks{}}

\date{}

\usepackage{graphicx}

\usepackage{geometry}
\begin{document}
\maketitle
\begin{abstract}

In today's society, location-based services are widely used which collect a huge amount of human trajectories.   Analyzing semantic meanings of these trajectories  can benefit numerous real-world applications, such as product advertisement, friend recommendation, and social behavior analysis. However, existing works on semantic trajectories are mostly centralized approaches that are not able to keep up with the rapidly growing trajectory collections. In this paper, we propose a novel large-scale semantic trajectory analysis algorithm in Apache Spark. We design a new hash function along with  efficient distributed algorithms that can quickly compute semantic trajectory similarities and identify communities of people with similar behavior across the world.  The experimental results show that our approach is more than 30 times faster than centralized approaches without sacrificing any accuracy like other parallel approaches.

\end{abstract}

\renewcommand\thesection{\Roman{section}}%

\section{Introduction}

Today, mobile devices with the Internet and positioning technology are extremely common, and almost all users of these devices utilize some forms of location-based services, such as using Google Maps to find nearby restaurants, getting directions from current locations,  and checking in on Yelp or other social network sites. As a result, a huge amount of trajectories has been collected by service providers.

\begin{figure*}[!t]
\centering
\includegraphics[width=6in]{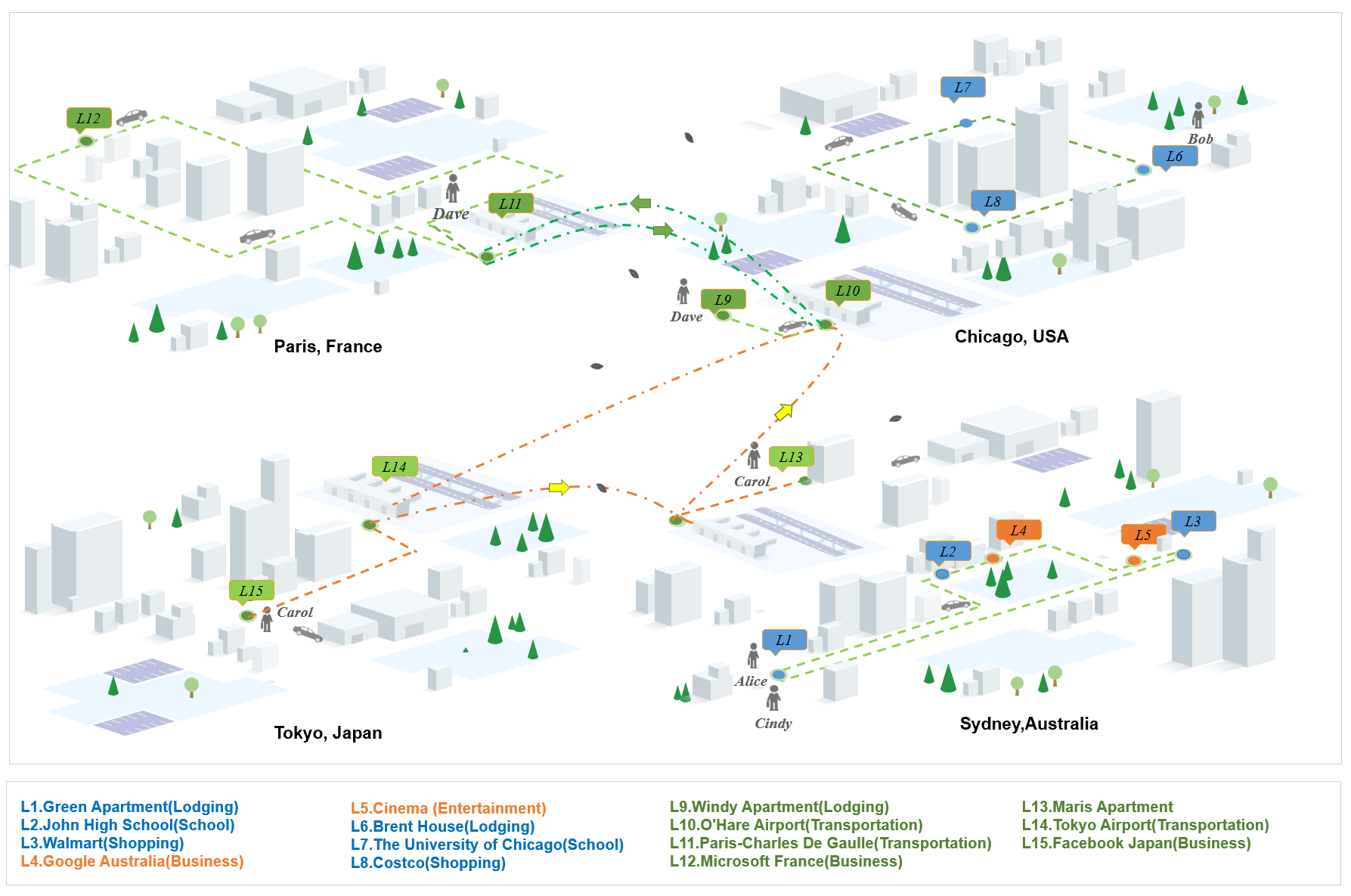}
\caption{Geometrically Similar Trajectories vs. Semantically Similar Trajectories: (1) Alice's and Cindy's trajectories are geometrically similar but semantically different; (2) Alice's and Bob's trajectories are geometrically different but semantically similar; (3) Dave and Carol also have geometrically different but semantically similar  trajectories as they are both frequent flyers. }
\label{fig:1}
\end{figure*}

Given the collected trajectories, people have looked into their  space similarities and semantic similarities. Identifying geometrically similar trajectories is beneficial for applications such as traffic flow analysis and urban planning, while analyzing semantic meanings of visited places may lead to even more interesting findings regarding humans' daily activity patterns and preferences.   For example, as shown in Figure \ref{fig:1}, Alice's and Cindy's trajectories in Sydney are overlapping, i.e.,  geometrically similar. However, we will find that they might have quite different interests after considering the semantic meanings of their trajectories. Specifically, Alice is visiting a high school and a grocery store while along the same path Cindy stops by  a  company and a cinema which demonstrates a different social behavior compared to Alice. In contrast, Alice's and Bob's trajectories have nothing in common in terms of spatial similarity as they live in different cities, whereas the semantic meanings of their trajectories reveal that they are both related with schools and share common daily activity patterns since they visited places of similar types. Similarly,  Carol and Dave demonstrate activity patterns of being frequent flyers although their trajectories are located in different cities and different countries.

The above examples illustrate the potential benefits of identifying semantically similar trajectories,  such as for targeted and precise product and service recommendations,  and online community formations. { Especially with the emerging of the future metaverse -- a social networking cyberspace where users have ``second" life through virtual reality and augmented reality technology, the boundaries between the physical world and cyber world are diminishing. Communities can be formed beyond those who are only spatially closed. Semantically similar trajectories in the real world could become a good indicator for friend exploration and recommendation in the metaverse. } However, it is indeed a very challenging task to analyze today's enormous volume of trajectories in a timely manner. The most representative and sophisticated techniques provide only centralized algorithms for calculating semantic similarity among trajectories \cite{xiao2010finding, xiao2014inferring}. It could take days and months for a centralized algorithm to analyze tens of billions of collected trajectories considering Facebook already has 2.4 billion active users. Moreover, unlike calculating the spatial similarity of trajectories which can be easily scaled up by parallelly processing trajectory datasets in different regions (e.g., cities) \cite{yuan2019distributed}, semantically similar trajectories are no longer confined by regions as they may be located in different countries as shown in the example. Thus, it is not a trivial task to design a divide-and-conquer strategy for semantic trajectory similarity calculation.

In order to address the aforementioned challenges, in this paper, we propose an efficient distributed computing algorithm on the Apache Spark platform \cite{zaharia2010spark} for analyzing large-scale semantic trajectories. Our algorithm is  named {\em AnotherMe} with the goal of finding people with similar social behavior across the world, who may share common interests and new insights. Specifically, we design a novel hash function, namely {\em Sequence-Sensitive Hashing (SSH)}, to effectively produce initial partitions of trajectories that are likely to be semantically similar. Then, we calculate trajectory similarity scores at different granularity levels in each partition in parallel. It is worth noting that our similarity function is more sophisticated than those in the previous works. Specifically, we take into account the recurrence of the same places in a trajectory (e.g.,company$\rightarrow$cafe$\rightarrow$company) which is commonly seen in the real world scenarios, whereas existing works simply consider distinct places in a trajectory and ignore their repetitive patterns.


Our proposed work has made the following contributions:
\begin{itemize}

\item Unlike most existing works on trajectory analysis \cite{cao2005mining, giannotti2007trajectory, yuan2019distributed} which mainly calculate the spatial similarity of trajectories, our work not only analyzes the more challenging type of similarity -- semantic similarity -- in a large scale setting,  but also provides a new way of identifying communities of common interests based on dynamic trajectory information rather than users' static profiles.

\item We design a novel hash function, the Sequence-Sensitive Hashing (SSH) which captures the characteristics of semantic similarity among trajectories much better than the original hashing function in the Apache Spark. Specifically, the trajectory analysis results obtained from our distributed algorithm have achieved 100\%  accuracy compared to the ground truth obtained from a centralized approach, whereas the Spark's original hash function yields very low accuracy.


\item We have an important finding regarding the implementation of distributed algorithms in Apache Spark. We found that although the common approach for implementing custom data analysis functions using Apache Spark's user-defined function feature is convenient, such an approach does not fully unleash Apache Spark's parallel computing capabilities. Significantly better performance is achieved after we strategically decompose and convert the tasks in the trajectory analysis process into atomic tasks that can be carried out directly by Apache Spark's built-in functions.

\item We  conduct extensive experiments on both real trajectory datasets and large-scale synthetic trajectory datasets. The results  prove that both the time and memory consumption of our method are  orders of magnitude less than the centralized version. { Specifically, our Spark-version is more than 30 times faster than the centralized approach}

\end{itemize}

The remainder of the paper is organized as follows. Section II provides a brief overview of the background knowledge about Apache Spark. Section III reviews the related work. Section IV presents our distributed trajectory analysis algorithms. Section V reports the experimental results. Finally, Section VI concludes the paper.

{
\section{Background of Apache Spark}

We first briefly introduce the key features of Apache Spark \cite{zaharia2010spark} as our algorithms are run on this environment. Spark is an open-source tool designed for large-scale data processing. It aims to address the performance limitations in an earlier popular large-scale data processing paradigm -- MapReduce by leveraging in-memory caching and processing. The fundamental data structure in Spark is  the resilient distributed dataset (RDD) which  helps parallelize distributed data processing over a cluster of machines. Spark also has an important component called Spark SQL  that supports parallel SQL operations which are used in our work. Spark SQL utilizes a data abstraction called DataFrames which can store structured and semi-structured data. Spark SQL can perform efficient SQL functions such as "group by" and "join". Moreover, Spark also provides several built-in hash functions for general uses, such as Bucketed Random Projection and MinHash. Overall, Apache Spark is the leading platform for the large-scale data analytics.
}

\section{Related Work}

In this section, we review related works on trajectory analysis. We will first quickly discuss early efforts that developed  centralized approaches for trajectory analysis. Then, we will analyze a few recent distributed approaches for large-scale trajectory datasets that are more related to our work.

{
Trajectory analysis is typically based on two kinds of similarities: coordinate-based similarity and semantic-based similarity.  The coordinate-based similarity measures include Euclidean distance \cite{faloutsos1994fast}, Hausdorff and Frechet distance, Angular metric or shape similarity \cite{chen2007spade,alt2009computational,nakamura2013shape} and edit distance on movement pattern strings\cite{chen2004symbolic}. The most commonly used coordinate-based similarity metric is Euclidean distance which aggregates the Euclidean distances of nearest pairs of positions in the two trajectories.  The semantics-based trajectory similarity is commonly measured using longest common subsequence or longest common sequence \cite{vlachos2002discovering, ying2010mining}. The longer the common subsequence, the less semantic-distance between the two trajectories, which means the more similar the two trajectories will be. The following is a simple example of the calculation of the coordinate-based similarity and the semantic-based similarity of two trajectories.  Assume Alice and Bob live in two different cities and their trajectories are represented as the place names and the coordinates as follows. Without loss of generality, we use integer coordinates for the ease of illustration.

$T_{Alice}$ = [high\_school(3,4) $\rightarrow$ shopping\_mall(5, 8) $\rightarrow$ dinner(10,7) $\rightarrow$ home(10,8)].

$T_{Bob}$ = [university(100,100) $\rightarrow$ grocery\_store(105, 110) $\rightarrow$ cafe(110,120) $\rightarrow$ home(120,125)].

The coordinate-based similarity between these two trajectories can be simply calculated by finding the average distance between each closet pair of locations, which is

\noindent{\small D=$\frac{1}{4}[$$\sqrt{(100-3)^2+(100-4)^2}$ + $\sqrt{(105-5)^2+(110-8)^2}$ + $\sqrt{(110-10)^2+(120-7)^2}$ + $\sqrt{(120-10)^2+(125-8)^2}]$=148}

This coordinate-based distance is very large due to the fact that the two people live in different cities. However, if we look at the semantic-based similarity between these two trajectories, we will notice that the places that Alice and Bob have visited belong to the same type and in the same sequence. Therefore, its semantic-based distance is close to 0. From here, we can observe that coordinate-based similarity is good at describing the spatial similarity of two trajectories while the semantic-based similarity is good at describing the social meaning of two trajectories.

\begin{figure*}[!t]
\centering{
\includegraphics[scale=0.60]{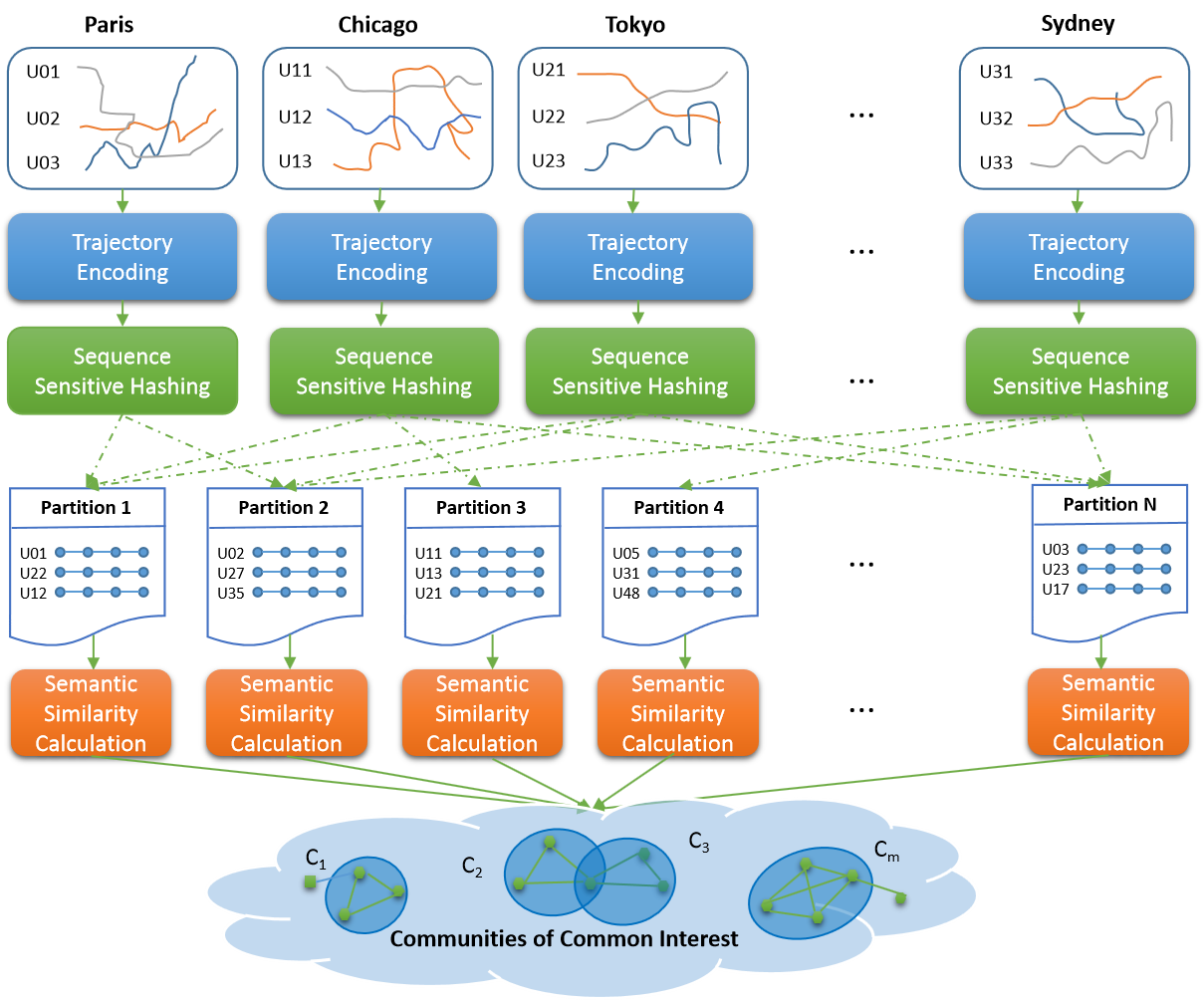}}
\caption{An Overview of the AnotherMe System}
\label{fig:4}
\end{figure*}

Various centralized approaches have been proposed to identify similar trajectories based on coordinate-based distance  \cite{cao2005mining,giannotti2007trajectory, luo2013finding, wang2017answering,10.1109/TPDS.2016.2565480}. Most of the early works focus on trajectory clustering \cite{jensen2007continuous} using Euclidean distances between coordinates of locations in trajectories. Later,  Zheng et al. \cite{zheng2008learning, zheng2008understanding, zheng2010understanding} started examining trajectory similarity in terms of semantic meanings of places visited.  Zhang  et al.~\cite{zhang2014splitter} also propose an approach namely SPLITTER to detect sequential patterns in semantic trajectories. The SPLITTER groups similar places together to mine coarse patterns,  and then progressively refine them using weighted snippet shifting to obtain fine-grained patterns. Choi et al.~\cite{choi2017efficient} devise a mining algorithm, called RegMiner, to find the local frequent movement patterns. Celik and Dokuz~\cite{celik2018discovering} further consider the temporal context of locations and only frequently visited or socially important locations when determining the similarity of trajectories.  Wan et al.~\cite{wan2017semantic} take into account both coordinate-based distance and semantic-based distance, and introduce a  new concept of semantic-geographic similar trajectories. {  Gao et al. \cite{GAO2020176} use clustering approaches to identify regions of interests and then represent trajectories in multi-resolutions. This multi-resolution idea is similar to our proposed hierarchical trajectory representation whereas we leverage the semantic hierarchy instead of the regions of interests. Most recently, semantic trajectory analysis has fostered a series of new applications, such as recommender systems in cultural spaces proposed by Angelis et al. \cite{bdcc5040080}, the itinerary recommender system by Cai et al. \cite{cai2018itinerary}, similar pattern group identification by Cao et al. \cite{cao2020effective}, and  social-space keyword query by \cite{CAO2021340}. } All of these works are centralized approaches and would not scale up when the huge amount of trajectories across the world need to be compared as that in our problem setting.

There have been some works on distributed trajectory analysis \cite{ choi2017efficient,chen2020parallel,yuan2019distributed}. However, most of them \cite{qin2019dfthr,zhang2017trajspark,ding2018ultraman, yuan2019distributed} use coordinate-based distance for trajectory comparison,  which eases the development of the parallel computation since trajectories can be easily divided into subsets based on regions they are located. Among these coordinate-based trajectory analyses works, many \cite{zhang2014efficient, eldawy2015spatialhadoop, xie2016simba} utilize MapReduce; very few \cite{zhang2017trajspark, ding2018ultraman} takes advantage of Apache Spark which has been shown to be much more efficient than MapReduce. A representative work using Apache Spark is TrajSpark \cite{zhang2017trajspark} which can answer k-nearest neighbor queries among trajectories based on the distance between coordinates of locations in the trajectories. Alvares et al. \cite{alvares2007model, alvares2007towards} append the semantic information to trajectories as a post-processing of the coordinate-based trajectory classification. The most recent work on distributed semantic trajectory analysis is by Chen et al. \cite{chen2020parallel}. Their approach first groups trajectories based on semantic meanings and then further partitions the trajectories based on spatial proximity. Unlike their work which leverages spatial partitioning to obtain a smaller dataset to process in parallel, the problem addressed in this paper aims to identify trajectories across the globe with similar semantic meanings which is much more challenging since we cannot employ spatial partitioning to reduce the number of trajectories that need an in-depth comparison.

To sum up, to the best of our knowledge, there has not been any distributed Spark-based algorithm for identifying semantic similarity among large-scale trajectories without spatial constraints.

\section{The AnotherMe Algorithm}
We propose a distributed  computing algorithm, called {\em AnotherMe}, which  analyzes a set of semantic trajectories and returns a set of communities of interest. The set of semantic trajectories belongs to people who may be in  different regions of the world. The communities of interest are formed by people who share common activity patterns as illustrated in the examples shown in Figure 1. The similarity between people's activity patterns is measured based on the similarity between their semantic trajectories. The formal definitions of these notions will be given in the following subsections.

Figure \ref{fig:4} presents an overview of our algorithm which consists of  four main phases: (i) Semantic encoding which converts the original semantic trajectory into an encoding of place types at different levels of semantic meanings; (ii) Sequence-Sensitive Hashing (SSH) which creates initial partitions of trajectories based on semantic similarities; (iii) Semantic trajectory similarity computation which calculates semantic trajectory similarity in each partition in parallel; and (iv) Output  communities of common interests, i.e., trajectories with same hash values and their pair-wise trajectory similarity.  In what follows, we elaborate the detailed algorithm in each phase.

\begin{figure}[h!]
\centering
\includegraphics[scale=0.37]{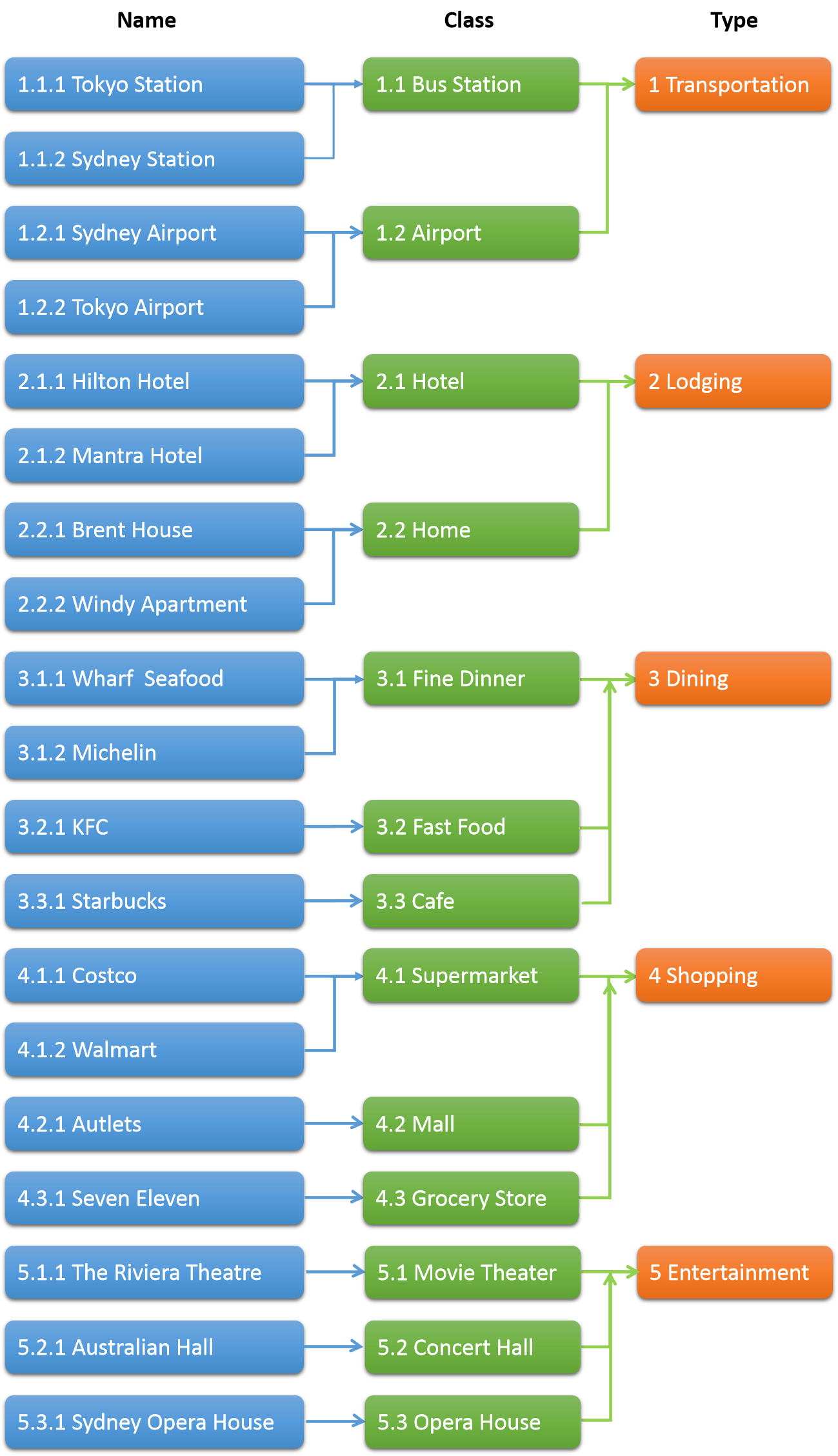}
\caption{An Example of Semantic Forest}
\label{fig:5}
\end{figure}

\subsection{Semantic Encoding}

The goal of this phase is to be able to generate an encoding for each trajectory so that we can easily  compare semantic trajectories (defined in Definition 1) that do not share any geographic overlap.

\begin{defn} {\em (Semantic Trajectory) A semantic trajectory of a user $u$ is in the form of $T_u$=[$place_1$ $\rightarrow$  $place_2$ $\rightarrow$ ... $\rightarrow$ $place_n$], where $place_i$ { ($1\leq i \leq n$)} denotes the name of the place where user $u$ stayed for more than $\tau$ time (a time parameter used to distinguish passing points from stay points), and the arrows indicate the visiting order.}
\end{defn}

The definition of the semantic trajectory considers mainly stay points,  which helps  the subsequent trajectory comparison to focus on meaningful places that a person has visited rather than just passed by briefly. Our representation capture the information of the length of stay by including the same place $n$ times if the duration of the stay at the place is $n$ times of the time interval $\tau$. This trajectory representation integrates spatial and temporal information that makes our proposed hash function more effective. By extending the  examples in Figure 1, the semantic trajectories of Carol and Dave can be represented as follows:

$T_{Carol}$ = [Maris Apartment $\rightarrow$ Sydney Airport $\rightarrow$ O'Hare Airport $\rightarrow$ Tokyo Airport $\rightarrow$ Facebook Japan$\rightarrow$ KFC $\rightarrow$ Tokyo Airport $\rightarrow$  Sydney Airport $\rightarrow$ Maris Apartment].

$T_{Dave}$ = [Windy Apartment $\rightarrow$ O'Hare Airport $\rightarrow$ Paris-Charles De Gaulle  $\rightarrow$ Microsoft France $\rightarrow$ Restaurant Goude $\rightarrow$ Paris Convention Center $\rightarrow$ Paris-Charles De Gaulle $\rightarrow$ O'Hare Airport $\rightarrow$ Windy Apartment].


From the above example, we can observe that Carol and Dave have a similar travel pattern between hotels and airports. However, such similarity cannot be identified if we simply try to match the place names in their trajectories. This is because they are visiting hotels and airports with different names that do not match directly. Therefore, besides the exact name matching, it is also important to consider the types of places. For this, we build a {\em semantic forest} to organize places into  three levels from fine granularity to coarse granularity, i.e., place name, place class, and place type.  Figure \ref{fig:5} shows an example of the semantic forest, which can be generated based on many existing ontology analysis of semantic meanings of words such as WordNet.

Each place in the semantic trajectory will be mapped to ``class" and ``type" levels in the semantic forest.  For example, at the ``class" level, Carol's trajectory will be represented as ``(hotel, airport, airport, airport, company, fast food, company, hotel)" , while Dave's trajectory will be represented as ``(home, airport, airport, company, fine dinner, convention, airport, airport, home)"; at ``type" level, Carol's trajectory will be  ``(lodging, transportation, transportation, transportation, business, dining, business, lodging)" while Dave's trajectory will be ``(lodging, transportation, transportation, business, dining, business, transportation, transportation, lodging)".

As a result, a single trajectory will have total three levels of representations: the original trajectory, the trajectory representation at ``class" level and the trajectory representation at ``type" level. The semantic similarity between a pair of trajectories is then defined as the integration of the similarity among three levels as shown in Definition 2.

{
\begin{defn}{\em (Semantic Trajectory Similarity) Given a pair of semantic trajectories $T_{u_1}$ and $T_{u_2}$, the semantic trajectory similarity between them is calculated as follows:

$SS(T_{u_1}, T_{u_2})$ = $\sum_{i=1}^n [w_i \cdot S(T^i_{u_1}, T^i_{u_2})]$,

where $T^i_{u_1}$ and $T^i_{u_2}$ represents the trajectory mapping at level $i$ among $n$ levels in the semantic forest, $w_i$ is the corresponding weight and $\sum_{i=1}^n w_i=1$, and $S(T^i_{u_1}, T^i_{u_2})$ denote the similarity of trajectories at level $i$.
}
\end{defn}
The above definition is generic that applies to any number of levels that a semantic forest may have. In our discussion, we will use 3 levels for ease of illustration. The definitions of the similarity among each level will be elaborated in Section IV.3.
}

Next, we propose a novel approach to consolidate the three-level representation of a trajectory to make the subsequent trajectory comparison highly efficient. Specifically, we map each place to an encoding in the form of ``$E_{type}$.$E_{class}$.$E_{name}$". For example, ``Tokyo station" is encoded as ``1.1.1", ``Sydney station" is encoded as ``1.1.2", and Sydney airport is encoded as ``1.2.1". When looking at the entire encodings of the above three places, we can see that they are different since the place names are different. However, when only the first two components of the encodings are considered, we can observe the similarity between ``Tokyo station" and ``Sydney station" as their encodings of the first two components are the same, i.e., ``1.1.x".  Finally, since all the three places have the same first component ``1.x.x" in their encodings, we also know that they belong to the same type. With the aid of the encoding, we will not need to store multiple representations for each trajectory. We only need to map the original trajectory to place encodings, and then the trajectory comparison at different levels can be conducted between  the extracted corresponding level of encodings  as shown in the above example. { Note that this encoding strategy can be easily extended to $n$-level representations of a trajectory in the form of ``$E_{l_1}.E_{l_2}...E_{l_n}$". }

\subsection{Sequence-Sensitive Hashing (SSH)}

Given a large amount of trajectory data, comparing each pair of trajectories to identify their similarity will be extremely time-consuming. For example, if there are 1 million (10$^6$) trajectories, there will be 1 trillion (10$^{12}$) pair-wise comparisons. To obtain a reasonable processing time, a natural idea is to adopt the divide-and-conquer method. This is easy when looking for the geometric similarity between trajectories whereby a large trajectory dataset can be split according to their regions (e.g., city or state), and then comparisons can be done within much smaller subsets of trajectories in parallel. As for semantic trajectories, spatial partitioning will not work since semantically similar trajectories could locate in different areas. The MinHashLSH function provided by Apache Spark does not work well either since MinHashLSH does not consider the frequency of the same type of places being visited. Thus, the partitioning of the semantic trajectories calls for a new strategy.

To conquer the aforementioned challenge, we propose a new hashing method, referred to as Sequence-Sensitive Hashing (SSH), which captures the semantic similarity among trajectories to help divide trajectories into semantic groups through only one scan of the trajectory dataset. Each semantic group contains trajectories that have similar types of places and are likely to have high similarity scores. Unlike geometric partitioning which does not have overlaps among partitions, our proposed semantic groups allow a trajectory to be classified as more than one semantic group. Such overlapping conforms with the real-world situation that a person's trajectory may demonstrate multiple characteristics, e.g., frequent flyer and businessman as shown in the previous example. Formally, our SSH method is built upon a new notion called {\em K-sequential Shingles} as defined in Definition \ref{def:shingle}.

\begin{defn} \label{def:shingle} {\em (K-sequential Shingling) {
Given a trajectory $T$=[$place_1$ $\rightarrow$  $place_2$ $\rightarrow$ ... $\rightarrow$ $place_n$], its k-sequential Shingling is a set of $m$ distinct shingles $G = \{g_1,g_2,...,g_m\}$, where each shingle is formed by the types of $k$ contiguous or non-contiguous places  in $T$ in the same order as they occur in the trajectory, i.e., $g_i$=[type$_{i_1}$ $\rightarrow$type$_{i_2}$...$\rightarrow$type$_{i_k}$], and $1\leq i \leq m$,  $m\leq {n \choose k}$.}}
\end{defn}
\vspace{3pt}

Figure \ref{fig:shingle} illustrates an example of 3-sequential shingles. The first two rows in the figure are the encoded trajectories of Carol and Dave, and the bottom two rows are their corresponding shingles formed by the types of places they have visited. { Specifically, the types of places (i.e., the first encoding of a place) visited by Carol include: 2 $\rightarrow$1$\rightarrow$1$\rightarrow$1$\rightarrow$6$\rightarrow$3$\rightarrow$6$\rightarrow$2, as highlighted using different colors in $T_{Carol}$.  The corresponding 3-shingles for the types of places are generated by selecting three place types from the trajectory while preserving their visiting sequence. For example, by selecting the 1st, 2nd and 3rd place types from the trajectory, we obtain the first 3-shingle "2 1 1" as denoted by orange and green colors, respectively; by selecting the 1st, 2nd and 5th place types, we obtain another 3-shingle "2 1 6". After collecting all the unique 3-shingles, we will obtain $S_{carol}$ as shown in Figure \ref{fig:shingle}. }

\begin{figure}[!ht]
\centering
\includegraphics[scale=0.45]{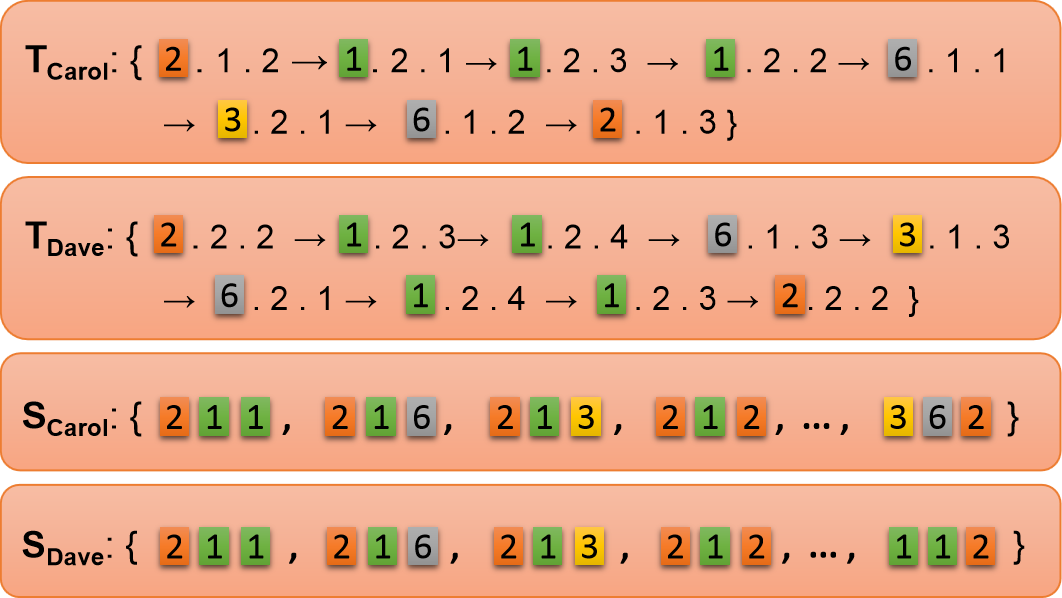}
\caption{An Example of 3-sequential Shingling}
\label{fig:shingle}
\end{figure}

\begin{figure*}[!t]
\centering
\includegraphics[scale=0.5]{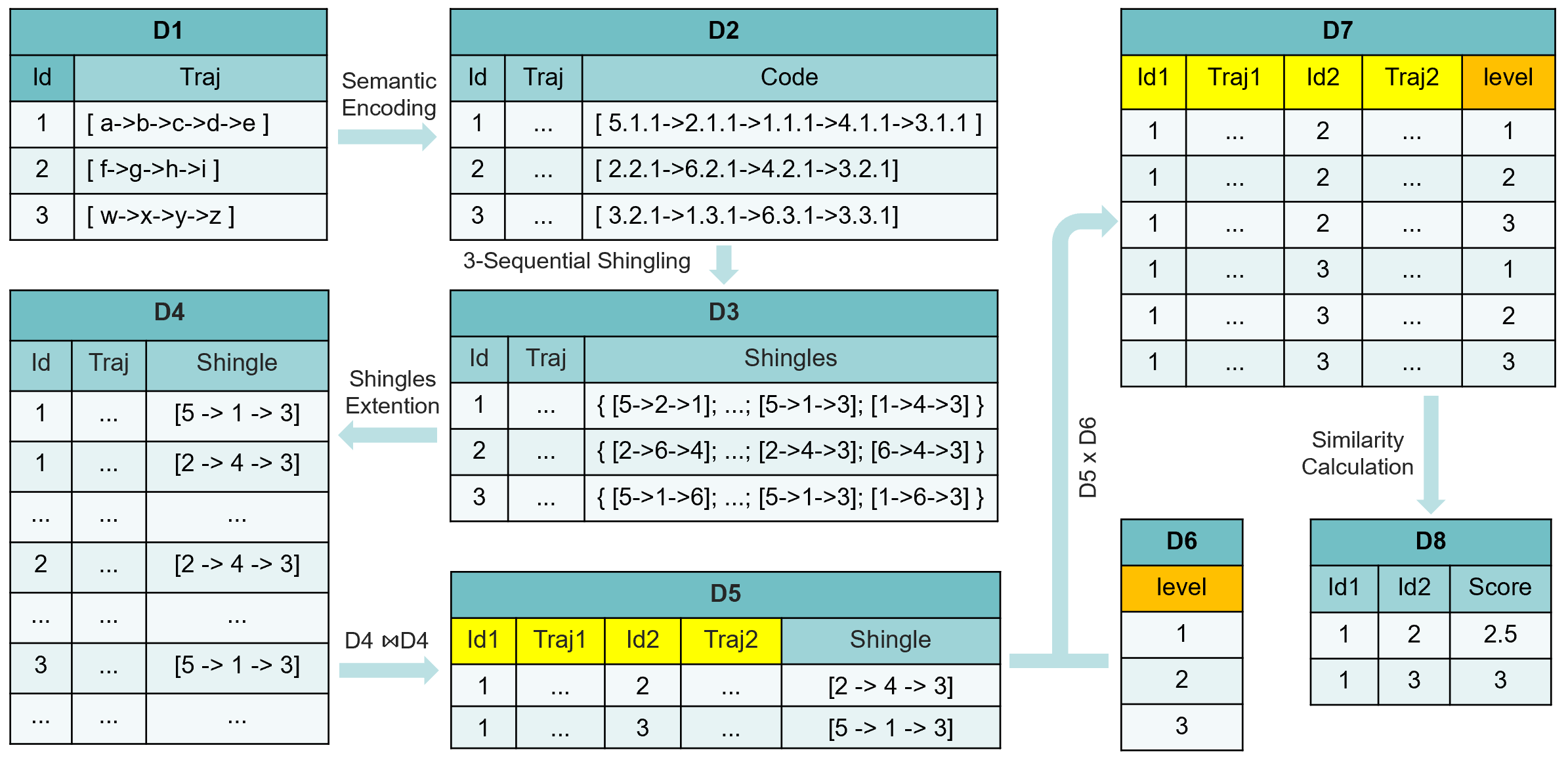}
\caption{ An Overview of Hashing Process}
\label{fig:6}
\end{figure*}
\begin{figure*}[ht]
\end{figure*}

Observing the shingles in Figure \ref{fig:shingle}, we can easily spot the similarities between the trajectories. In order to capture such similarities, we calculate the k-sequential shingles for each trajectory and then cluster the trajectories which contain at least one identical shingle into the same partition (semantic group).  Note that, if two trajectories have not even a single common k-sequential shingle, that means the two trajectories did not visit more than $k$ same types of places in the same order, and hence they are not very similar and it is not necessary to further calculate their similarity score.


We proceed to present how to implement the Sequence-Sensitive Hashing (SSH) in Apache Spark. A straightforward method is to implement SSH as a user-defined function (UDF) in Apache Spark. However, we found that the use of UDF does not fully take advantage of Apache Spark's parallel computing capabilities due to the limitation of Apache Spark's abilities to understand arbitrary UDF codes. Also, it is hard for UDF to benefit from the Apache Spark query optimizer called Catalyst which keeps improving Apache Spark's built-in functions with each new version release. As such, we design the following algorithm that leverages Apache Spark's built-in SQL query functions to conduct the proposed trajectory partitioning task.

Figure \ref{fig:6} shows an example of hashing process. The original trajectories are stored as a tuple of $\langle ID, T_{ID}\rangle$ in a data frame (denoted as $D_1$), where $ID$ is a unique ID for each trajectory $T_{ID}$. For each trajectory, its semantic encoding is appended as a new column in $D_2$. Based on the semantic encoding, we  calculate the k-sequential shingles for each trajectory and store the results in the data frame $D_3$. In $D_3$, each trajectory may contain various numbers of shingles. We further expand $D_3$ into a new data frame $D_4$ by listing each trajectory shingle as a separate row.  Then,  we can leverage the Apache Spark join function to join the $D_4$ with itself to identify trajectories containing identical shingles. The join results are stored in another data frame ($D_5$). Each row of $D_5$ contains a pair of the IDs of trajectories that are likely to be similar to each other.

\begin{algorithm}[!t]
  \caption{ 3-sequential Shingling}
  \KwIn{A trajectory containing a sequence of location: $T = x_1,x_2,x_3,...,x_l$}
  \KwOut{A set of shingles: $S = \{s_1,s_2,...,s_z\}$ }
  \For{$i=1;i \le l;i++$}
  {
    \For{$j=i+1;j \le l;j++$}
    {
        \For{$k=j+1;j \le l;k++$}
        {
            s=concat($x_i$.name$\|$$x_j$.name$\|$$x_k$.name);
            S.insert(s);
        }
    }
  }
  return $S$\;
\end{algorithm}

{

We now take a deep look at the SSH.
 Given Algorithm 1, let $N$ be the number of trajectories and $L$ be the average number of locations in a trajectory. The computational complexity of Sequential-Sensitive Hashing is $O(N*L^3)$. Since the total number of trajectories is much larger than the average number of locations in a trajectory, i.e., $N>>L$, the actual computational complexity of SSH is $O(N)$ which is linear to the data size.

Next, we examine  the collision rate of the SSH. In our problem, the SSH is used to help group similar trajectories. Therefore, we do not want the collision rate to be too small. For example, if the collision rate is 0, it means each trajectory is in its own group and we will not be able to find any similar trajectories. Assume there are $Q$ items in total in the type level of semantic forest. We will obtain at most $Q^K$ different hash values by applying K-sequential Shingling for all. Each trajectory will have ${L \choose K}$ shingles. On average, the collision rate is ${L \choose K}/Q^K$.
}

\subsection{Semantic Trajectory Similarity Computation}

\begin{figure*}[!ht]
\centering
\includegraphics[scale=0.60]{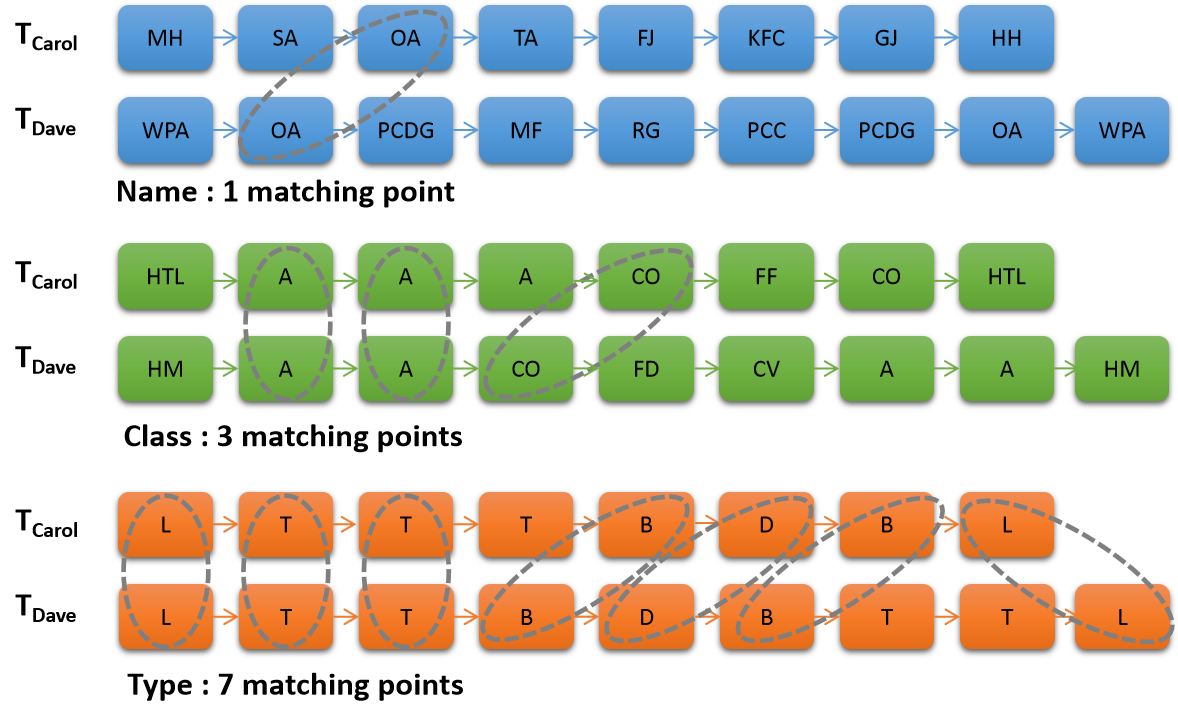}
\caption{An Example of Multi-level Semantic Similarity}
\label{fig:10}
\end{figure*}
\par
\begin{figure*}
\begin{equation}
M_h(ET_u^h, ET_v^h)=\left\{
  \begin{array}{ll}
\emptyset, &~~~if~~i=0~~or~~j=0; \\
M_h(ET_{u_{i-1}}^h, ET_{v_{j-1}}^h)\bigoplus h_{u_i}, &~~~if~~h_{u_i}=h_{v_j}; \\
max\{M_h(ET_{u_{i-1}}^h,ET_{vj}^h), M_h(ET_{u_i}^h,ET_{v_{j-1}}^h)\},&~~~others.
  \end{array}
\right.\label{equ:mmes}
\end{equation}
\end{figure*}

After the trajectory partitioning, we  only need to compare similarities among trajectories within the same partition. In order to better capture the semantic similarity of trajectories that contain repetitive visits to the same type of places as discussed in the introduction, we introduce a new definition of semantic trajectory similarity as follows.

\vspace{5pt}
{
\begin{defn} {\em (Multi-Level Semantic Trajectory Similarity) Let $ET_{u}$ = \{$E_{u_1}^{l_1}.E_{u_1}^{l_2}...E_{u_1}^{l_n}$, ...,  $E_{u_k}^{l_1}.E_{u_k}^{l_2}...E_{u_k}^{l_n}$ \},
$ET_{v}$=\{$E_{v_1}^{l_1}.E_{v_1}^{l_2}...E_{v_1}^{l_n}$, ...,  $E_{v_m}^{l_1}.E_{v_m}^{l_2}...E_{v_m}^{l_n}$ \},
denote the semantic encodings of user $u$'s and $v$'s trajectories where $k$ and $m$ refer to the number of places visited by each user, respectively. The multi-level semantic trajectory similarity (MSS) is defined as:

\vspace{5pt}

\noindent MSS($ET_{u}$, $ET_{v}$) = $\sum_{h=1}^{n}\beta_{h}\cdot|M_h|$

\vspace{5pt}

\noindent where $M_h$ contains the maximum matching encoding sequence in two trajectories at level $h$,
$\beta_h$ denotes the importance of the similarity at level $l_i$, and $\sum_{i=1}^{n}\beta_{l_i}=1$.
}
\end{defn}
\vspace{5pt}

}

\begin{algorithm}[!t]
  \caption{{Sequence Sensitive Hashing}}
  \KwIn{A Dataframe containing all semantic trajectories: $D_{1}=\{T_1, T_2,...,T_N\}$}
  \KwOut{A Dataframe containing all pairs of candidate similar trajectories: $D_{5}$}\
  Encode the place names of each trajectory in Dataframe $D_1$ into corresponding place types in semantic trees, get Dataframe $D_2$\;
  Use 3-Sequential Shingling function to calculate a set of shingles of each trajectory in Dataframe $D_2$ and add a new column named ``shingles" to store it, get Dataframe $D_3$\;
  Extend the Dataframe $D_3$ into a new Dataframe $D_4$ by separating the elements of shingles set of every trajectory into different rows\;
  Join Dataframe $D_4$ with itself on condition:  left $D_4$.id != right $D_4$.id and left $D_4$.shingles = right $D_4$.shingles, get Dataframe $D_5$\;
  return $d_{5}$\;
\end{algorithm}

Considering the three-level semantic hierarchy, two points that match at type level do not necessarily match at class or place level. However, two points matching at class level must match at type level and two points matching at place level must match at class level and type level. Therefore, $|M_{typ}| \geq |M_{cls}| \geq |M_{p}|$. In addition, $\beta_{typ}+\beta_{cls}+\beta{p}=1$, so given a similarity threshold $n \leq \rho < n+1$, where $n$ is a positive integer denoting the lower bound of $\rho$. Trajectories with a similarity score $MSS > \rho$ has $|M_{typ}|$ matching points at type level, and $n \leq MSS \leq |M_{typ}|$. In this way, our Sequential Sensitive Hashing applying k-sequential shingle function at type level with a $k \leq n+1$ is able to cluster all the possible similar pairs of trajectories satisfying the condition above into partitions, and it performs most efficiently when $k = n+1$.

\vspace{5pt}

\begin{defn} {\em (Maximum Matching Encoding Sequence) Let $ET_u^h$ = \{$h_{u_1}$, $h_{u_2}$, ..., $h_{u_k}$\}, $ET_w^h$=\{$h_{v_1}$, $h_{v_2}$, ..., $h_{v_m}$\} denote the semantic encodings of user $u$'s and $v$'s trajectories at level h. Let $M_h$($ET_{u_i}^h$, $ET_{v_j}^h$) represents the longest common subsequence of prefixes till $h_u^i$ and $h_v^j$ in the two trajectories,  where $0\leq i\leq k$, $0\leq j\leq m$, and $k$ and $m$ refer to the number of places visited by the two users. The maximum matching encoding sequence is defined in Equation \ref{equ:mmes}.}
\end{defn}

\vspace{5pt}

Figure \ref{fig:10} shows an example of the multi-level semantic trajectory similarity calculation.  For  convenience, place names, classes, and types are represented using their abbreviations. In this example, the number of matching points increases as the semantic meaning becomes more general. Specifically, there is only one matching place in Carol's and Dave's trajectories, but three places with matching classes, and seven places with matching types. Thus,  if given the importance of the similarity at different level as $\beta_{typ}=0.2$, $\beta_{cls}=0.3$, and $\beta_{p}=0.5$, the similarity score MSS($ET_{Carol}$, $ET_{Dave}$) = $\beta_{typ}\cdot7$ + $\beta_{cls}\cdot3$ + $\beta_{p}\cdot1$ = 0.2$\times$7+0.3$\times$3+0.5$\times$1=2.8. { Here, the weight values $\beta$ are chosen in a descending order of their importance whereby the highest weight value is given to the lowest semantic level, i.e., the place level.}

It is worth noting that our algorithm is capable of capturing the repetition of visits to the same places. However, existing definitions of semantic trajectory similarity will consider the same place only once \cite{xiao2010finding, li2008mining} which will not be able to distinguish frequent visits from occasional visits, such as frequent flyers vs. occasional travelers, customers to a shopping mall vs. sale persons in a shopping mall.


As shown in Figure \ref{fig:6}, the calculation of the similarity score is realized as follows in the Spark. We create a new data frame  $D_6$ that has one column named "level", and three rows with values: 1, 2, and 3 which represent the "type", "class", and "name" level in the semantic forest respectively. { We then multiple  $D_5$ with $D_6$ to obtain a new data frame $D_7$. In $D_7$, each row is a concatenation of two trajectories and the semantic level ID (e.g., 1, 2, or 3) at which we want to calculate the similarity score for these two trajectories. This multiplication helps separate the similarity calculation at different semantic levels, so the parallel computing capability of Spark can be better utilized when  we calculate the semantic similarity score for each row in $D_7$}. We append a new column "score" to $D_7$ and obtain a new data frame $D_8$. The last step is to apply the Apache Spark's aggregation function to sum the scores for each pair of $id_1$ and $id_2$, which yields the final trajectory similarity score.

In addition, the same pairs of trajectories may occur in multiple partitions. For example, Carol's and Dave's trajectories may appear together in two partitions: one is for frequent flyers and the other is for businessmen. In this case, our approach will ensure that the similarity between the same pair of trajectories will be calculated only once. Specifically, each pair of the trajectories will be loaded to the main memory only once no matter how many partitions they belong to. This is achieved by joining two candidate  trajectories that share any identical shingles, but no matter how many shingles they share. Our approach  can save a lot of redundant calculations on duplicate pairs.

%
%

{
\subsection{Algorithm Outline}

We now summarize the AnotherMe algorithm. Given a set of semantic trajectories, our algorithm starts from converting each trajectory into its corresponding  semantic encoding. Based on the semantic encoding, we further calculate the k-sequential shingles for each trajectory. Then, we group trajectories which have at least one identical shingle into the same partition. Up to this point, we obtain multiple data partitions, and these partitions may contain overlapping trajectories since one person may belong to multiple social groups.  Finally, inside each  partition, we leverage the Spark built-in functions to calculate the similarity scores of each pair of trajectories. The obtained similarity scores can be further used to identify communities of interests, each of which contains a group of trajectories that are similar to one another.
}

\section{Experimental Studies}

In this section, we first introduce experimental settings and then report the comparison results of our approach against centralized approaches and the default hashing function in Apache Spark.

\subsection{Experimental Settings}

We have implemented our AnotherMe algorithm in  Spark. The testing environment consists of  1 master node and 1 to 20 worker nodes. Each node has 8 CPUs with Intel Xeon Cascade Lake 8255C (2.5 GHz), 32G of memory and 1T disk space available. The number of executor in Spark is the same as the number of worker nodes.



We compare  our proposed AnotherMe algorithm with the following four algorithms.
\begin{itemize}
\item Centralized: This is a centralized approach that performs the calculation on the entire dataset. It calculates the semantic similarity of each pair of trajectories.


\item MinHash: This is a distributed approach that employs the MinHashLSH function provided by Apache Spark to generate hash values for each trajectory and group trajectories with similar hash values for further similarity calculation using the function defined in Definition 4. Specifically, it first builds a vocabulary of place types and then encodes each trajectory at the type level into a single binary vector. In the binary vector,  non-zero values are used to represent the presence of place type in the corresponding trajectory. The MinHashLSH is applied to the trajectory encodings to obtain hash values.


\item BRP: This is also a distributed approach that employs another hash function, namely Bucketed Random Projection (BRP), provided by Apache Spark.


\item { User-defined: We implemented another version of our algorithm by enclosing all the tasks as a stand-alone user-defined function in Apache Spark.}  This algorithm has the same logic (i.e., the same four phases) as AnotherMe with the only difference in implementation. Recall that AnotherMe breaks down the tasks into subtasks that can leverage the built-in functions such as group-by in Apache Spark.

\end{itemize}

We test the performance of different approaches using both synthetic and real datasets. The synthetic dataset contains up to 1 million trajectories. The length of each trajectory is defined as the number of semantic locations visited, which varies from 5 to 10 to mimic the average person's daily travel patterns. Each location on the trajectories is randomly selected from 10,000 places. The number of  synthetic place type is 30 and the number of classes in each type is 10. We set the importance of the similarity at different level as $\beta_{typ}=\beta_{cls}=\beta{p}=1/3$. Besides synthetic datasets, there are not any real datasets of semantic trajectories. The largest real trajectory dataset we can find is  GeoLife~\cite{zheng2010geolife, zheng2008understanding, zheng2009mining}, which contains 17,621 trajectories collected from 182 users by GPS devices. The trajectories in GeoLife are a sequence of GPS locations. In order to carry out semantic queries, we first detect stay points in the given GPS trajectories \cite{li2008mining, xiao2010finding} and then map the stay points to semantic names. Finally, we assign each converted semantic trajectory a user ID.

By default, we apply 3-Sequential Shingling in our SSH algorithm which means that trajectories with less than three matching locations will not be  considered for further analysis. The reason to use at least 3-sequential shingling is due to the consideration that most of the trajectories already contain at least two common types of places, i.e., home and workplace. Two similar trajectories are expected to visit at least one more additional type of place. In terms of the final query result, we only report to the query issuer those trajectories with a similarity score greater than threshold 2.


The performance is evaluated using two major criteria: (i) CPU time; (ii) accuracy. The accuracy is defined by comparing the query output obtained from the distributed approaches with that obtained from the centralized approach. The centralized approach calculates the similarity scores among all pairs of trajectories,  based on which it  constructs maximal cliques that connect users whose trajectories' similarity scores are above a threshold to form communities  of interests. The communities of interests identified by the centralized approach is used as the ground truth to evaluate the accuracy of distributed approaches. The distributed approaches will split the dataset into subsets and then calculate the similarity among trajectories in each subset. If the splitting strategy is not well designed, it is likely that some similar trajectories are classified into different subsets and missed in the query result. Formally, the query accuracy is defined as follows:
\begin{equation}
    QA_1=\frac{|N_{dis}\bigcap N_{cen}|}{|N_{cen}|}
\end{equation}
In the equation, the query accuracy $QA_1$ is defined as the number of matching communities of common interests obtained from the distributed and centralized approach divided by the number of communities of common interests found by the centralized approach ($N_{cen}$). { The use of $QA_1$ aims to match how the real-world applications could use the semantically similar trajectories for friends or social circle recommendations. Besides this $QA_1$ metric, we also adopt another direct metric $QA_2$  that simply compares the total number of similar trajectory pairs identified by the centralized approach and distributed approaches. Equation \ref{eq:qa2} provides the formal definition of $QA_2$, whereby $ST_{dis}$ and $ST_{cen}$ denote the number of similar trajectory pairs output by the distributed and centralized approaches, respectively.
\begin{equation} \label{eq:qa2}
    QA_2=\frac{|ST_{dis}\bigcap ST_{cen}|}{|ST_{cen}|}
\end{equation}
}

\subsection{Comparison with the Centralized Approach}

In the first round of experiments, we compare the four approaches by varying the number of trajectories from 10,000 to 60,000. Note that 60,000 is the maximum number of trajectories that the centralized approach can be tested in a single computing node without encountering the problem of memory explosion and requiring hours for a single run.

Figure \ref{fig:16} shows the overall processing time of the five algorithms in a single node: Centralized, MinHash, BRP, User-defined, and AnotherMe. The processing time taken by Centralized and User-defined approaches increases exponentially with the data size. They are extremely slower  than MinHash, BRP, and AnotherMe. It is not surprising to see that the centralized approach requires extensive time to calculate the similarity scores between each pair of trajectories in the dataset whereas other approaches only need to compare smaller sets of candidate trajectories that are likely to be similar. At a closer look at these approaches, Figure \ref{fig:15} shows the  actual number of trajectories being compared by each approach, and Figure \ref{fig:hashtime} shows the portion of time taken by hash functions in each distributed approach. We can observe that the AnotherMe approach is slightly slower than MinHash and BRP when the dataset size increases. This is not because the computation efficiency of AnotherMe decreases. This is because MinHash and BRP calculate smaller number of similar trajectories  since these two hash functions cannot accurately find out all similar trajectory pairs, and hence seem to be faster. The accuracy of these approaches are compared in Figure \ref{fig:17}  as discussed later.

\begin{figure}[!ht]
\centering
\includegraphics[scale=0.62]{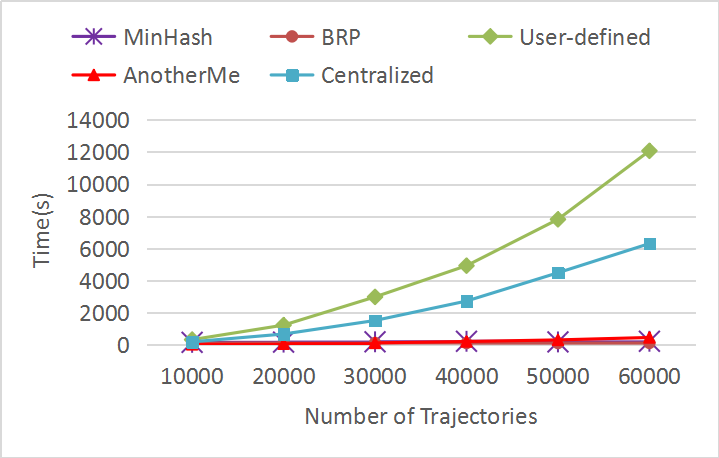}
\caption{Overall Processing Time when Varying the Number of Trajectories}
\label{fig:16}
\end{figure}

\begin{figure}[!ht]
\centering
\includegraphics[scale=0.65]{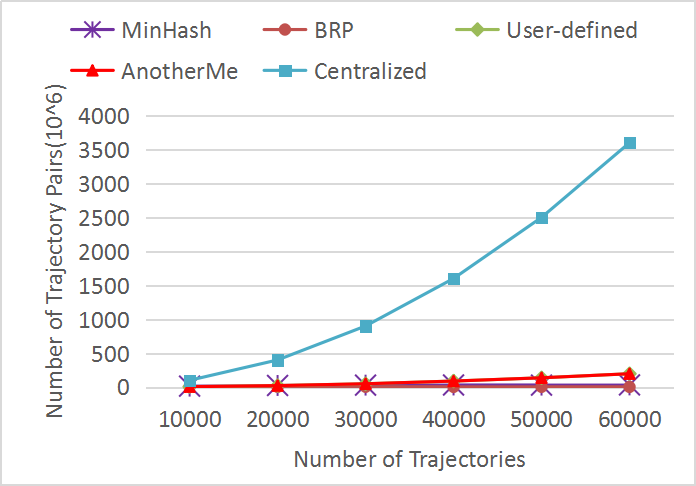}
\caption{ Number of Trajectory Pairs Being Compared when Varying the Number of Trajectories}
\label{fig:15}
\end{figure}

\begin{figure}[h!]
\centering
\includegraphics[scale=0.6]{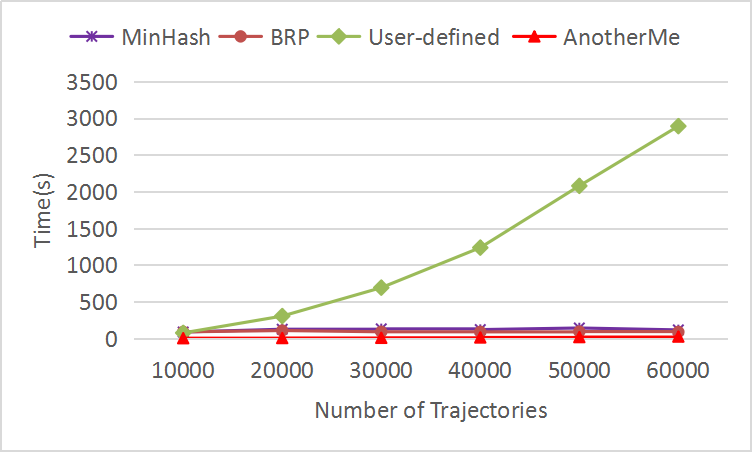}
\caption{Time Overhead Incurred by Hash Functions}
\label{fig:hashtime}
\end{figure}

An interesting phenomenon in this experiment is that the User-defined approach in Apache Spark becomes even slower than the centralized approach when the dataset size increases.  Note that each test dataset fits in the main memory during the comparison, which rules out the potential cause of the performance degradation of the User-defined approach due to the delay of data being swapped between memory and disk. We may have expected that a distributed approach would be generally faster than a centralized approach. However, our new finding is that Apache Spark simply treats user-defined functions as a black box and is not able to optimize and fully parallelize the processes in the user-defined functions. Also, due to the extra environmental setting overhead in Spark, the user-defined function approach becomes less efficient than the lightweight centralized approach.

From the experiments, we also observe that our proposed AnotherMe approach is not only very fast but also yields the highest accuracy (e.g., 100\%). As shown in Figure \ref{fig:17}, { the AnotherMe approach identifies  both the same number of communities of interests (measured by $QA_1$) and the same number of similar trajectories (measured by $QA_2$) as the centralized approach. This means the AnotherMe's hash function  did not split similar trajectories into different data partitions. The user-defined approach also has high accuracy as it is another implementation of the  same logic as AnotherMe. The results clearly indicate that our proposed Sequence-Sensitive Hashing (SSH) very well captures the similarity among trajectories. In contrast, the accuracy of the approaches using the default hash functions (MinHash and BRP) in Apache Spark drops dramatically with the increase of the number of trajectories. This is mainly because that the default Apache Spark hash functions are not tailored for the proposed trajectory similarity comparison.  The default hash functions do not capture the visiting order of different places, and hence it is very likely that the default hash functions generate similar hash values for trajectories which are indeed different because of the order of places visited. As a result, the accuracy of the MinHash and BRP is very low with BRP missing almost all the correct communities and similar trajectory pairs. }

\begin{figure}[!t]
\centering
\includegraphics[scale=0.6]{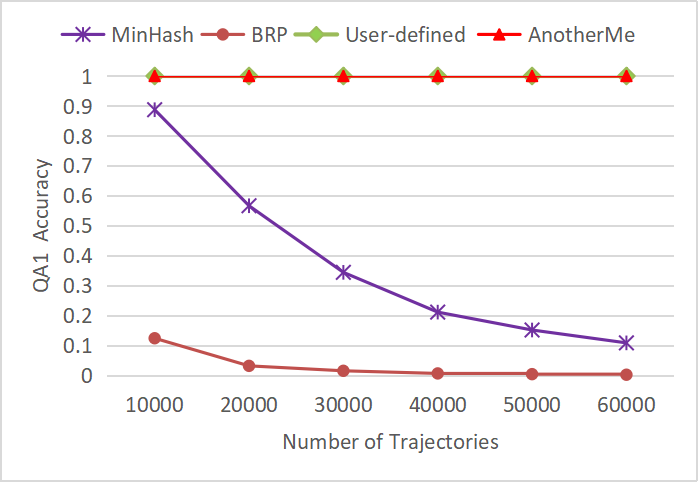}
\includegraphics[scale=0.6]{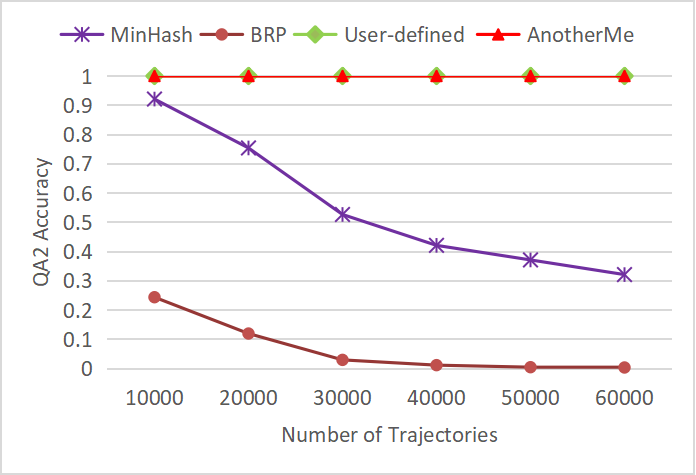}
\caption{Accuracy When Varying the Number of Trajectories}
\label{fig:17}
\end{figure}

\subsection{Evaluating Real Dataset}

In this round of experiments, we examine the performance in the real dataset, the GeoLife as described in the experimental settings. We compare all the approaches except the BRP which is not able to correctly detect most of the communities of interests. Figure \ref{fig:13} shows the overall processing time. As we can see from the figure, our proposed AnotherMe approach is the most efficient, faster than MinHash,  and  10 times faster than the centralized approach. This indicates the practicality of our approach in the real world. In addition, we also observe that the user-defined approach is faster than the centralized approach in this case. This is mainly because the size of the real dataset is not very large, and hence not many computing nodes are needed for the user-defined approach, i.e., less communication overhead.

\begin{figure}[!ht]
\centering
\includegraphics[scale=0.6]{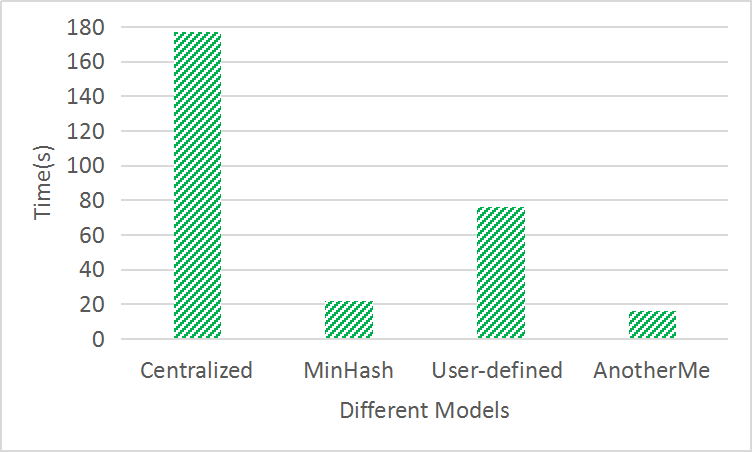}
\caption{Overall Processing Time of Different Approaches in Real Datasets}
\label{fig:13}
\end{figure}

Next, we look at the accuracy of all the approaches. As shown in Figure \ref{fig:14}, our proposed AnotherMe and its implementation as the user-defined function both achieve 100\% accuracy compared to the centralized approach in terms of both metrics $QA_1$ and $QA_2$. However, the default Apache Spark hash function, the MinHash, drops the accuracy to around 80\%. The reason is that MinHash was developed as a general hashing function for different applications by Apache Spark. AnotherMe has a carefully crafted SSH hashing for trajectory similarity detection and hence achieves high accuracy.

Considering both efficiency and accuracy altogether, the proposed AnotherMe is superior to all the other approaches.


\begin{figure}[!t]
\centering
\includegraphics[scale=0.6]{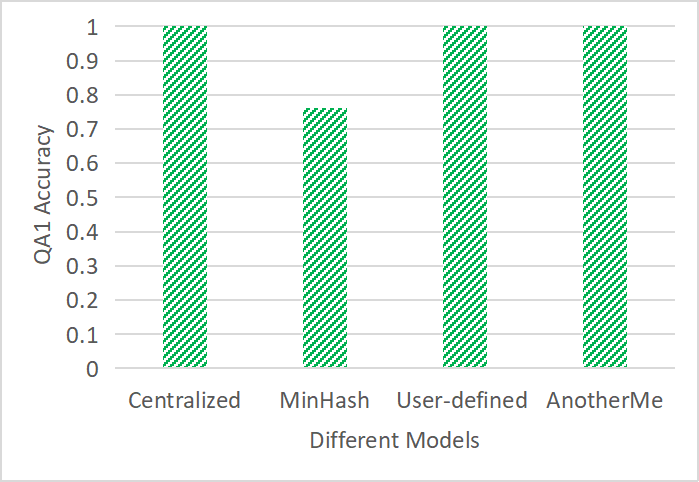}
\includegraphics[scale=0.6]{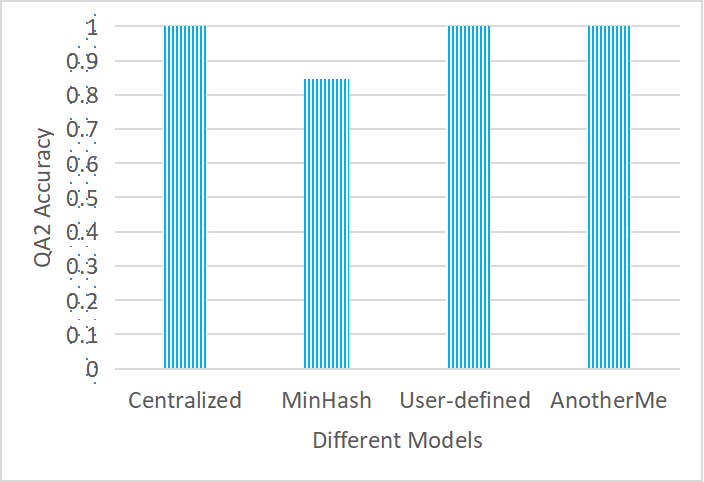}
\caption{Accuracy of Different Approaches in Real Datasets}
\label{fig:14}
\end{figure}

\subsection{Evaluating Scalability}

In this round of experiments, we aim to examine the scalability of the distributed approaches. We only compare the MinHash and  our proposed AnotherMe approaches, since the centralized approach and user-defined function approach are no longer able to process large-scale datasets within hours, and the BRP approach has extremely low accuracy. The  number of trajectories being tested has been increased up to 1 million and the number of different place types is set to 300.


Figure \ref{fig:18} reports the overall processing time of the MinHash and our AnotherMe algorithm when 4 worker nodes are used.    We can clearly see that our approach is orders of magnitude faster than MinHash. The performance gain achieved by our AnotherMe approach should be attributed to the essential design spirit of AnotherMe which carefully maps the subtasks to highly-optimized Apache Spark's built-in operations on data frames. Moreover, since the number of place types is larger than that tested in smaller dataset, the benefits of our hashing algorithm in terms of capturing similar trajectory pairs and balancing the workload among nodes become more prominent.

\begin{figure}[!t]
\centering
\includegraphics[scale=0.6]{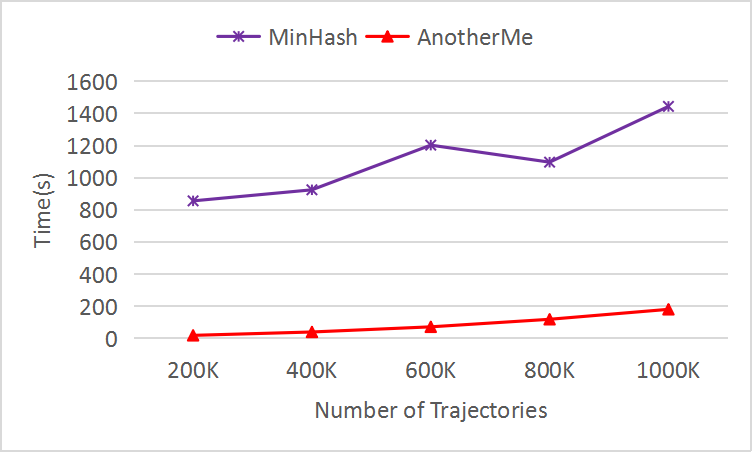}
\caption{Processing Time when Varying the Data Size}
\label{fig:18}
\end{figure}
\begin{figure}[!t]
\centering
\includegraphics[scale=0.6]{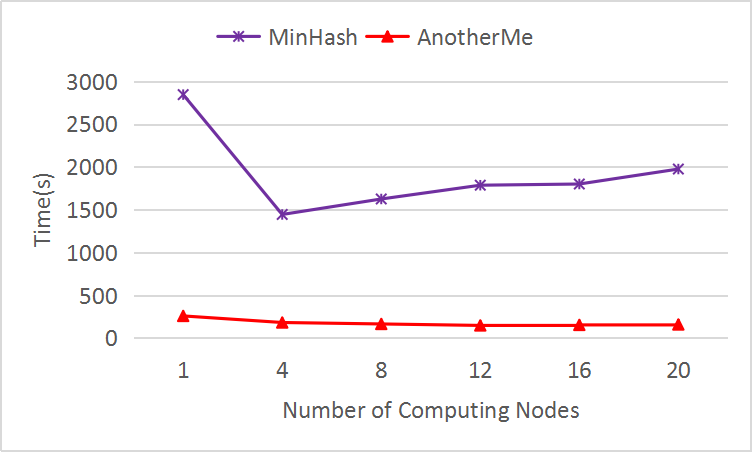}
\caption{Processing Time  when Varying the Number of Computing Nodes}
\label{fig:19}
\end{figure}

In Figure \ref{fig:19}, we vary the number of worker nodes from 1 to 20 and test the  dataset of 1 million  trajectories. Both approaches speed up with the increase of the number of worker nodes at first and then the performance does not further improve much. This is because of the effect of two factors. The more worker nodes, the fewer number of calculations to be conducted in individual nodes. However, more worker nodes also introduce more communication overhead such as data shuffling among nodes, which in turn introduces additional processing time. In our experiments, MinHash reaches its  best performance when 4 worker nodes are employed, while AnotherMe reaches its best performance when 12 worker nodes are employed.

\begin{figure}[!t]
\centering
\includegraphics[scale=0.6]{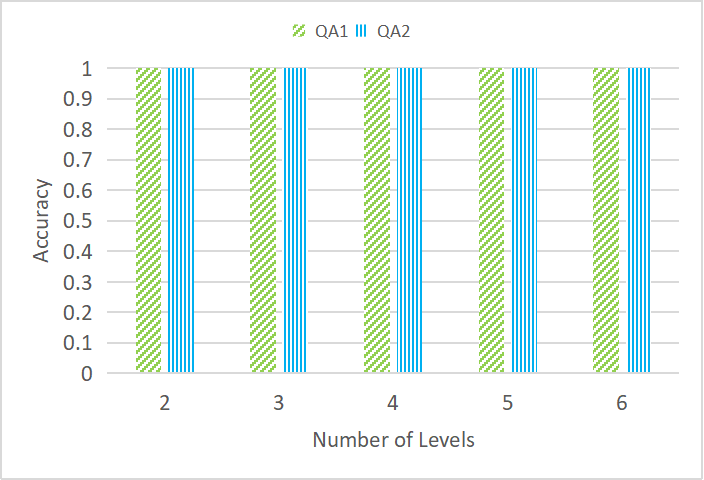}
\includegraphics[scale=0.6]{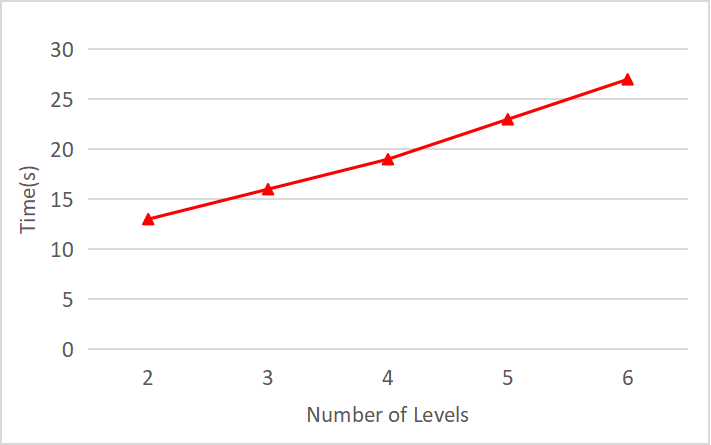}
\caption{Effect of the Number of Semantic Levels}
\label{exp:qalevel}
\end{figure}

{
\subsection{Evaluating Effect of Semantic Levels}

In the previous experiments, we have adopted the 3-level  semantic hierarchy. In this last round of experiments, we are interested in investigating the generality of our algorithm in future applications which may use different numbers of semantic levels. The experiments are conducted on the real trajectory dataset (GeoLife), whereby we adopt semantic hierarchy with levels ranging from 2 to 6. The accuracy is measured by comparing the number of communities and the number of similar trajectory pairs found by our algorithm with the centralized approach. Figure \ref{exp:qalevel} shows the accuracy of our algorithm in terms of the metrics $QA_1$ and $QA_2$. As we can see that our algorithm performs consistently regardless the number of semantic levels. Specifically, when similarity measures are defined over different number of semantic levels, our algorithm still identifies the same similar trajectory pairs and communities as the centralized approach. This indicates certain robustness and generality of our algorithm. In addition, we also evaluate the processing time of our algorithm under these different settings. It is not surprising to see an increase of the time when the number of semantic levels grows. This is because our hashing needs to be performed at each level of the semantic hierarchy, and hence more levels will need more time to process.
}

\section{Conclusion}

In this paper,  we propose an efficient large-scale semantic trajectory analysis algorithm (dubbed AnotherMe) on the Apache Spark platform. Specifically, we design a novel hash function to effectively produce initial partitions of trajectories that are likely to be semantically similar. Then, we calculate trajectory similarity scores at different semantic granularity levels in each partition in parallel, which significantly improves the comparison speed. The experimental results show that our approach is orders of magnitude faster than centralized approaches while preserving the same accuracy. The obtained similarity scores between trajectories are beneficial for identifying communities of common interests across the physical boundaries. { In addition, our proposed multi-level similarity metric and sequence-sensitive hashing technique may also be useful for other applications that need sequence analysis. For example, our approach may be leveraged to  analyze the multiple steps that malware is taking to perform infection and identify similar malware behavior. The potential of the proposed technique is worth exploring in various application domains. }


\bibliographystyle{unsrt}
\bibliography{references}

\begin{thebibliography}{10}

\bibitem{xiao2010finding}
Xiangye Xiao, Yu~Zheng, Qiong Luo, and Xing Xie.
\newblock Finding similar users using category-based location history.
\newblock In {\em Proceedings of the 18th SIGSPATIAL international conference
  on advances in geographic information systems}, pages 442--445, 2010.

\bibitem{xiao2014inferring}
Xiangye Xiao, Yu~Zheng, Qiong Luo, and Xing Xie.
\newblock Inferring social ties between users with human location history.
\newblock {\em Journal of Ambient Intelligence and Humanized Computing},
  5(1):3--19, 2014.

\bibitem{yuan2019distributed}
Haitao Yuan and Guoliang Li.
\newblock Distributed in-memory trajectory similarity search and join on road
  network.
\newblock In {\em 2019 IEEE 35th International Conference on Data Engineering
  (ICDE)}, pages 1262--1273. IEEE, 2019.

\bibitem{zaharia2010spark}
Matei Zaharia, Mosharaf Chowdhury, Michael~J Franklin, Scott Shenker, and Ion
  Stoica.
\newblock Spark: Cluster computing with working sets.
\newblock {\em HotCloud}, 10(10-10):95, 2010.

\bibitem{cao2005mining}
Huiping Cao, Nikos Mamoulis, and David~W Cheung.
\newblock Mining frequent spatio-temporal sequential patterns.
\newblock In {\em Fifth IEEE International Conference on Data Mining
  (ICDM'05)}, pages 8--pp. IEEE, 2005.

\bibitem{giannotti2007trajectory}
Fosca Giannotti, Mirco Nanni, Fabio Pinelli, and Dino Pedreschi.
\newblock Trajectory pattern mining.
\newblock In {\em Proceedings of the 13th ACM SIGKDD international conference
  on Knowledge discovery and data mining}, pages 330--339, 2007.

\bibitem{faloutsos1994fast}
Christos Faloutsos, Mudumbai Ranganathan, and Yannis Manolopoulos.
\newblock Fast subsequence matching in time-series databases.
\newblock {\em Acm Sigmod Record}, 23(2):419--429, 1994.

\bibitem{chen2007spade}
Yueguo Chen, Mario~A Nascimento, Beng~Chin Ooi, and Anthony~KH Tung.
\newblock Spade: On shape-based pattern detection in streaming time series.
\newblock In {\em 2007 IEEE 23rd International Conference on Data Engineering},
  pages 786--795. IEEE, 2007.

\bibitem{alt2009computational}
Helmut Alt.
\newblock The computational geometry of comparing shapes.
\newblock In {\em Efficient Algorithms}, pages 235--248. Springer, 2009.

\bibitem{nakamura2013shape}
Tetsuya Nakamura, Keishi Taki, Hiroki Nomiya, Kazuhiro Seki, and Kuniaki
  Uehara.
\newblock A shape-based similarity measure for time series data with ensemble
  learning.
\newblock {\em Pattern Analysis and Applications}, 16(4):535--548, 2013.

\bibitem{chen2004symbolic}
Lei Chen, M~Tamer {\"O}zsu, and Vincent Oria.
\newblock Symbolic representation and retrieval of moving object trajectories.
\newblock In {\em Proceedings of the 6th ACM SIGMM international workshop on
  Multimedia information retrieval}, pages 227--234, 2004.

\bibitem{vlachos2002discovering}
Michail Vlachos, George Kollios, and Dimitrios Gunopulos.
\newblock Discovering similar multidimensional trajectories.
\newblock In {\em Proceedings 18th international conference on data
  engineering}, pages 673--684. IEEE, 2002.

\bibitem{ying2010mining}
Josh Jia-Ching Ying, Eric Hsueh-Chan Lu, Wang-Chien Lee, Tz-Chiao Weng, and
  Vincent~S Tseng.
\newblock Mining user similarity from semantic trajectories.
\newblock In {\em Proceedings of the 2nd ACM SIGSPATIAL International Workshop
  on Location Based Social Networks}, pages 19--26, 2010.

\bibitem{luo2013finding}
Wuman Luo, Haoyu Tan, Lei Chen, and Lionel~M Ni.
\newblock Finding time period-based most frequent path in big trajectory data.
\newblock In {\em Proceedings of the 2013 ACM SIGMOD international conference
  on management of data}, pages 713--724, 2013.

\bibitem{wang2017answering}
Sheng Wang, Zhifeng Bao, J~Shane Culpepper, Timos Sellis, Mark Sanderson, and
  Xiaolin Qin.
\newblock Answering top-k exemplar trajectory queries.
\newblock In {\em 2017 IEEE 33rd International Conference on Data Engineering
  (ICDE)}, pages 597--608. IEEE, 2017.

\bibitem{10.1109/TPDS.2016.2565480}
Albino Altomare, Eugenio Cesario, Carmela Comito, Fabrizio Marozzo, and
  Domenico Talia.
\newblock Trajectory pattern mining for urban computing in the cloud.
\newblock {\em IEEE Transaction on Parallel and Distributed Systems (TPDS)},
  28(2):586--599, 2017.

\bibitem{jensen2007continuous}
Christian~S Jensen, Dan Lin, and Beng~Chin Ooi.
\newblock Continuous clustering of moving objects.
\newblock {\em IEEE transactions on knowledge and data engineering},
  19(9):1161--1174, 2007.

\bibitem{zheng2008learning}
Yu~Zheng, Like Liu, Longhao Wang, and Xing Xie.
\newblock Learning transportation mode from raw gps data for geographic
  applications on the web.
\newblock In {\em Proceedings of the 17th international conference on World
  Wide Web}, pages 247--256. ACM, 2008.

\bibitem{zheng2008understanding}
Yu~Zheng, Quannan Li, Yukun Chen, Xing Xie, and Wei-Ying Ma.
\newblock Understanding mobility based on gps data.
\newblock In {\em Proceedings of the 10th international conference on
  Ubiquitous computing}, pages 312--321. ACM, 2008.

\bibitem{zheng2010understanding}
Yu~Zheng, Yukun Chen, Quannan Li, Xing Xie, and Wei-Ying Ma.
\newblock Understanding transportation modes based on gps data for web
  applications.
\newblock {\em ACM Transactions on the Web (TWEB)}, 4(1):1, 2010.

\bibitem{zhang2014splitter}
Chao Zhang, Jiawei Han, Lidan Shou, Jiajun Lu, and Thomas La~Porta.
\newblock Splitter: Mining fine-grained sequential patterns in semantic
  trajectories.
\newblock {\em Proceedings of the VLDB}, 7(9):769--780, 2014.

\bibitem{choi2017efficient}
Dong-Wan Choi, Jian Pei, and Thomas Heinis.
\newblock Efficient mining of regional movement patterns in semantic
  trajectories.
\newblock {\em Proceedings of the VLDB Endowment}, 10(13):2073--2084, 2017.

\bibitem{celik2018discovering}
Mete Celik and Ahmet~Sakir Dokuz.
\newblock Discovering socially similar users in social media datasets based on
  their socially important locations.
\newblock {\em Information Processing \& Management}, 54(6):1154--1168, 2018.

\bibitem{wan2017semantic}
You Wan, Chenghu Zhou, and Tao Pei.
\newblock Semantic-geographic trajectory pattern mining based on a new
  similarity measurement.
\newblock {\em ISPRS International Journal of Geo-Information}, 6(7):212, 2017.

\bibitem{GAO2020176}
Chongming Gao, Zhong Zhang, Chen Huang, Hongzhi Yin, Qinli Yang, and Junming
  Shao.
\newblock Semantic trajectory representation and retrieval via hierarchical
  embedding.
\newblock {\em Information Sciences}, 538:176--192, 2020.

\bibitem{bdcc5040080}
Sotiris Angelis, Konstantinos Kotis, and Dimitris Spiliotopoulos.
\newblock Semantic trajectory analytics and recommender systems in cultural
  spaces.
\newblock {\em Big Data and Cognitive Computing}, 5(4), 2021.

\bibitem{cai2018itinerary}
Guochen Cai, Kyungmi Lee, and Ickjai Lee.
\newblock Itinerary recommender system with semantic trajectory pattern mining
  from geo-tagged photos.
\newblock {\em Expert Systems with Applications}, 94:32--40, 2018.

\bibitem{cao2020effective}
Yang Cao, Fei Xue, Yuanying Chi, Zhiming Ding, Limin Guo, Zhi Cai, and
  Hengliang Tang.
\newblock Effective spatio-temporal semantic trajectory generation for similar
  pattern group identification.
\newblock {\em International Journal of Machine Learning and Cybernetics},
  11(2):287--300, 2020.

\bibitem{CAO2021340}
Keyan Cao, Qimeng Sun, Haoli Liu, Yefan Liu, Gongjie Meng, and Jingjing Guo.
\newblock Social space keyword query based on semantic trajectory.
\newblock {\em Neurocomputing}, 428:340--351, 2021.

\bibitem{chen2020parallel}
Lisi Chen, Shuo Shang, Christian~S Jensen, Bin Yao, and Panos Kalnis.
\newblock Parallel semantic trajectory similarity join.
\newblock In {\em 2020 IEEE 36th International Conference on Data Engineering
  (ICDE)}, pages 997--1008. IEEE, 2020.

\bibitem{qin2019dfthr}
Jiwei Qin, Liangli Ma, and Qing Liu.
\newblock Dfthr: A distributed framework for trajectory similarity query based
  on hbase and redis.
\newblock {\em Information}, 10(2):77, 2019.

\bibitem{zhang2017trajspark}
Zhigang Zhang, Cheqing Jin, Jiali Mao, Xiaolin Yang, and Aoying Zhou.
\newblock Trajspark: A scalable and efficient in-memory management system for
  big trajectory data.
\newblock In {\em Asia-Pacific Web (APWeb) and Web-Age Information Management
  (WAIM) Joint Conference on Web and Big Data}, pages 11--26. Springer, 2017.

\bibitem{ding2018ultraman}
Xin Ding, Lu~Chen, Yunjun Gao, Christian~S Jensen, and Hujun Bao.
\newblock Ultraman: a unified platform for big trajectory data management and
  analytics.
\newblock {\em Proceedings of the VLDB Endowment}, 11(7):787--799, 2018.

\bibitem{zhang2014efficient}
Yu~Zhang, Youzhong Ma, and Xiaofeng Meng.
\newblock Efficient spatio-textual similarity join using mapreduce.
\newblock In {\em 2014 IEEE/WIC/ACM International Joint Conferences on Web
  Intelligence (WI) and Intelligent Agent Technologies (IAT)}, volume~1, pages
  52--59. IEEE, 2014.

\bibitem{eldawy2015spatialhadoop}
Ahmed Eldawy and Mohamed~F Mokbel.
\newblock Spatialhadoop: A mapreduce framework for spatial data.
\newblock In {\em 2015 IEEE 31st international conference on Data Engineering},
  pages 1352--1363. IEEE, 2015.

\bibitem{xie2016simba}
Dong Xie, Feifei Li, Bin Yao, Gefei Li, Liang Zhou, and Minyi Guo.
\newblock Simba: Efficient in-memory spatial analytics.
\newblock In {\em Proceedings of the 2016 International Conference on
  Management of Data}, pages 1071--1085, 2016.

\bibitem{alvares2007model}
Luis~Otavio Alvares, Vania Bogorny, Bart Kuijpers, Jose Antonio~Fernandes
  de~Macedo, Bart Moelans, and Alejandro Vaisman.
\newblock A model for enriching trajectories with semantic geographical
  information.
\newblock In {\em Proceedings of the 15th annual ACM international symposium on
  Advances in geographic information systems}, pages 1--8, 2007.

\bibitem{alvares2007towards}
Luis~Otavio Alvares, Vania Bogorny, Bart Kuijpers, Bart Moelans, Jose~Antonio
  Fern, ED~Macedo, and Andrey~Tietbohl Palma.
\newblock Towards semantic trajectory knowledge discovery.
\newblock {\em Data Mining and Knowledge Discovery}, 12, 2007.

\bibitem{li2008mining}
Quannan Li, Yu~Zheng, Xing Xie, Yukun Chen, Wenyu Liu, and Wei-Ying Ma.
\newblock Mining user similarity based on location history.
\newblock In {\em Proceedings of the 16th ACM SIGSPATIAL international
  conference on Advances in geographic information systems}, pages 1--10, 2008.

\bibitem{zheng2010geolife}
Yu~Zheng, Xing Xie, Wei-Ying Ma, et~al.
\newblock Geolife: A collaborative social networking service among user,
  location and trajectory.
\newblock {\em IEEE Data Eng. Bull.}, 33(2):32--39, 2010.

\bibitem{zheng2009mining}
Yu~Zheng, Lizhu Zhang, Xing Xie, and Wei-Ying Ma.
\newblock Mining interesting locations and travel sequences from gps
  trajectories.
\newblock In {\em Proceedings of the 18th international conference on World
  wide web}, pages 791--800, 2009.

\end{thebibliography}

\begin{IEEEbiography}[{\includegraphics[width=1in,height=1.25in,clip,keepaspectratio]{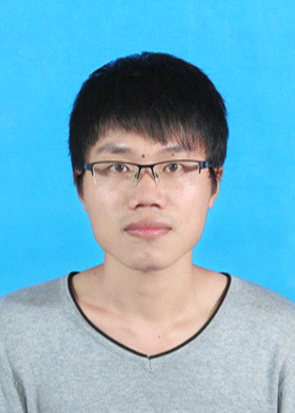}}]{Chaoquan Cai} received a BS degree in Software Engineering and an MS degree in Computer Science from Fuzhou University, Fuzhou, China. He is currently working towards a Computer Science PhD degree at University of Missouri. His research interests include statistical machine learning, large-scale semantic data management and analysis.
\end{IEEEbiography}
\vspace{-55 mm}
\begin{IEEEbiography}[{\includegraphics[width=1in,height=1.25in,clip,keepaspectratio]{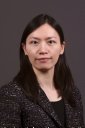}}]{Dan Lin} received a PhD degree in Computer Science from  National University of Singapore in 2007, and was a postdoctoral research associate at Purdue University for two years. She is currently an associate professor and Director of I-Privacy Lab at University of Missouri. Her research interests cover many areas in the fields of information security and database systems.
\end{IEEEbiography}

\end{document}